\documentclass[12pt]{article}
\usepackage{graphicx}
\usepackage{pdfpages}
\usepackage{adjustbox}
\usepackage{amsmath,amsfonts,amssymb,amsthm}
\usepackage{caption}
\usepackage[ruled,vlined]{algorithm2e}
\usepackage{bbm}
\usepackage{breakcites}
\usepackage{tabularx}
\usepackage{geometry}
\usepackage{enumitem}
\usepackage{natbib} 
\usepackage{authblk}
\usepackage{graphicx}
\usepackage{subcaption}
\usepackage{hyperref}
\usepackage{ulem}
\usepackage{mathtools}

\setlist[itemize]{align=parleft,left=0pt..1em}
\newcolumntype{L}{>{\raggedright\arraybackslash}X}

\usepackage{stix}
\usepackage{color}

\DeclareMathOperator*{\argmin}{argmin}

\theoremstyle{definition}
\newtheorem{definition}{Definition}[section]
\newtheorem{assumption}{Assumption}
\newtheorem{example}{Example}

\newtheorem{theorem}{Theorem}[section]

\newtheorem{lemma}[theorem]{Lemma}

\providecommand{\keywords}[1]
{
  \small	
  \textbf{{Keywords:}} #1
}

\makeatletter
\renewcommand\AB@affilsepx{, \protect\Affilfont}
\makeatother

     \title{Design-Based Inference for Multi-arm Bandits}
    
    \author[1]{Dae Woong Ham\thanks{Email: \href{mailto:daewoongham@g.harvard.edu}{\tt daewoongham@g.harvard.edu}}} 
    \author[2]{Iavor Bojinov}
    \author[3]{Michael Lindon}
    \author[3]{Martin Tingley}

    \affil[1]{\textit{Department of Statistics, Harvard University}}
    \affil[2]{\textit{Harvard School of Business}}
    \affil[3]{\textit{Netflix, U.S.A}}
  
    \date{   \vspace{-0.5cm} \today}
    
\begin{document}
    
    \maketitle
\vspace{-1cm}
\begin{abstract}
  Multi-arm bandits are gaining popularity as they enable real-world sequential decision-making across application areas, including clinical trials, recommender systems, and online decision-making. Consequently, there is an increased desire to use the available adaptively collected datasets to distinguish whether one arm was more effective than the other, \emph{e.g.}, which product or treatment was more effective. Unfortunately, existing tools fail to provide valid inference when data is collected adaptively or require many untestable and technical assumptions, \emph{e.g.}, stationarity, $iid$ rewards, bounded random variables, etc. Our paper introduces the design-based approach to inference for multi-arm bandits, where we condition the full set of potential outcomes and perform inference on the obtained sample. Our paper constructs valid confidence intervals for both the reward mean of any arm and the mean reward difference between any arms in an assumption-light manner, allowing the rewards to be arbitrarily distributed, non-$iid$, and from non-stationary distributions. In addition to confidence intervals, we also provide valid design-based confidence \textit{sequences}, sequences of confidence intervals that have uniform type-1 error guarantees over time. Confidence sequences allow the agent to perform a hypothesis test as the data arrives sequentially and stop the experiment as soon as the agent is satisfied with the inference, \emph{e.g.}, the mean reward of an arm is statistically significantly higher than a desired threshold. 
\end{abstract}


\newpage

\section{Introduction}


Multi-armed bandit (MAB) algorithms are a popular and well-established framework for sequential decision-making \cite{Robbins_MAB, Berry_bandit, nonstat_ex1, sutton_MAB, MAB_overview}. Because of their regret-minimizing properties, MABs are used across various applications such as optimal treatment allocation in clinical trials \cite{MAB_clinical_trials}, recommender systems improvements \cite{rec_MAB}, anomaly detection for networks \cite{anomaly_MAB}, and aiding online decision-making \cite{netflix_MAB}. For example, \citeauthor{MAB_clinical_trials} uses an adaptive bandit algorithm to sequentially assign treatments against skin cancer in a mice experiment. Here, the authors care both about assigning the best possible treatment and estimating the relative effectiveness of the different treatments. 


As the clinical trial example illustrates, real-world applications of MABs increasingly require inference (i.e., determining if one of the arms is significantly better than the alternatives). Unfortunately, inference is challenging because MAB algorithms collect data adaptively, breaking the standard independent and identically distributed ($iid$) assumption often evoked in the statistics literature. To make progress, researchers have focused on specific settings, such as linear bandits where the rewards follow a parametric linear model \citep{linear_ex3, linear_ex2, linear_ex1}, or have made strong untestable assumptions on both the rewards and the action space, such as stationarity and $iid$ rewards \citep{kelly_batched_bandits, hadad, ian_bandit}. 

Broadly, there are two approaches for inference in MAB problems. The first is to perform inference at the \textit{end} of a pre-specified time by constructing a single confidence \textit{interval} (CI). The second is to perform inference \textit{during} the study as new data arrives by constructing a confidence \textit{sequence} --- a sequence of confidence intervals that are uniformly valid over time. As we perform inference on every arm at every time, the second approach allows us to stop collecting the data once we detect that one of the arms is statistically significantly outperforming the others. 

In our paper, we propose a generic and flexible framework for performing inference both during and at the end of the study. Our work leverages the design-based approach to causal inference, which has a long history in the statistics community dating back to Fisher and Neyman but has seen a resurgence in popularity as it permits for inference on the obtained sample while handling complicated settings such as interference in an assumption-light manner \cite{fisher:1935, neym:35, rubin_design1, Holland_designbased, rubin:imbens, peng_luke_design, guillaume_design, mypaper}. The design-based framework has several advantages in MAB settings. First, it lets us relax the stationary and independent assumption for the rewards because the design-based framework conditions on the (potential) outcomes, allowing our inferential results to hold for any general reward distribution.\footnote{For example, the mean reward of each arm may be time-varying and also dependent on past data.} Second, a central principle in any MAB algorithm is balancing exploration versus exploitation; however, this trade-off is inherently about analyzing how the agent has performed for the \textit{current} sample, \emph{e.g.}, how much the agent ``lost'' by picking arm A over arm B for the finite-sample data. Unlike the more common super-population approach that performs inference on the general population eligible to interact with the MAB algorithm \cite{hadad, ian_bandit}, the design-based framework directly performs inference for relevant finite-sample estimands. Lastly, while relaxing the stationary and independence assumptions, the design-based framework only requires light and mostly testable assumptions. In particular, the main assumption we require is that the probability of picking an arm at any time is bounded away from zero or one. Since the agent/experimenter is in control of the data collection process, this can not only be verified but also controlled by using the appropriate adaptive scheme.

The main contribution of this paper is that we formally introduce the design-based approach for MAB, allowing us to build valid confidence intervals (Section~\ref{section:CI_results}) and confidence sequences (Section~\ref{section:CS_results}) for the mean reward of each arm and the difference between two arms' mean rewards. We define our setting and design-based estimand and estimators in Section~\ref{section:setting}. Then we extend our results to contextual MAB in Section~\ref{section:CMAB} and end with simulations in Section~\ref{section:sims}.

\section{Setup: Estimands and Estimation}
\label{section:setting}
We now introduce relevant notation, the design-based inference framework, the estimands of interest in the MAB framework, and our proposed estimators. 
\subsection{Design-Based Causal Inference}
\label{subsection:DB_intro}
Suppose we observe $T$ samples of $\{W_t, Y_t\}_{t = 1}^T$, where $W_t \in \{0, \dots, K-1\}$ is the $K$ possible actions (or treatments) the agent takes at time $t$ and $Y_t$ is the observed reward (or outcome) at time $t = 1,\dots, T$. For readers more familiar with the MAB literature, the notation $(A_t, R_t)$ is often used for the action and reward, respectively. We purposefully use the alternative notation to emphasize the connection with causal inference. In Section~\ref{section:CMAB}, we further assume that at each point in time, we also observe some covariates or contexts $X_t$ to generalize our analysis to contextual MAB \cite{CMAB_ref}.

Borrowing from the standard causal inference literature, we assume that for each $t$ there are $K$ potential outcomes $Y_t(0),\dots,Y_t(K-1)$, corresponding to the $K$ possible actions \cite{fisher:1935, neym:35, rubin:imbens,abadie2020sampling}. Note that we implicitly assume the no-interference assumption of potential outcomes  for simplicity and brevity of notation. To connect the potential outcomes to the observed outcome, we assume that $y_t = Y_t(w_t)$, where lower case $(w_t, y_t)$ denotes our observed treatment samples. 

The design-based approach conditions on the full set of potential rewards $Y_t(0),\dots,Y_t(K-1)$, allowing the rewards to be arbitrarily dependent on the past rewards and non-stationary. This setting generalizes the common assumptions invoked in MAB, where researchers assume $Y_t(W_t)$ is a function of only its current action and is generated from an $iid$ distribution. The relaxation of the assumptions is possible because the design-based approach shifts the modeling burden away from the unknown outcome to the know action distribution.

Specifically, we assume for each $t$ the action $W_t$ to be adapted based on the historical data,
\begin{equation}
p_{t \mid t-1}(w) := \Pr(W_t = w_t \mid \mathcal{F}_{t-1}),
\label{eq:adaptive_trt}
\end{equation}
where the filtration $\mathcal{F}_{t -1}$ contains the information set of the past $(W_{1:(t -1)}, Y_{1:(t-1)})$. In Section~\ref{section:CMAB}, we extend Equation~\eqref{eq:adaptive_trt} to allow the action to further depend on the context variables. Lastly, we impose that $p_{t \mid t-1}(w)$ is bounded away from zero or one, often known as probabilistic treatment assignment assumption in causal inference \cite{rubin:imbens}.
\begin{assumption}[Probabilistic Treatment Assignment]
\label{assumption:PTA}
For all times $t = 1, \dots, T$ and possible actions $w \in \{0, \dots, K-1\}$, we have that
$$0 < p_{t \mid t-1}(w) < 1.$$
\end{assumption}
Assumption~\ref{assumption:PTA} is fairly weak and holds as long as the proposed MAB algorithm does not converge to a zero/one probability event for any of the arms in finite time. Furthermore, we remark that many existing works require this regularity condition for inference \cite{hadad, kelly_lucas, mypaper}.

\subsection{Multi-arm Bandit Estimands}
\label{subsection:MAB_intro}
The primary estimand of interest in MAB algorithms is the reward mean distribution for arm $w\in \{0,\dots,K-1\}$. In the typical MAB literature, the reward mean distribution of arm $w$ is distributed $iid$, thus can be characterized through $\mu_t(w) = \mu(w)$ since it does not vary over time. As we do not impose any restriction on the reward mean, we choose to define our estimand as a finite-sample cumulative mean:
\begin{equation}
\label{eq:Q_func}
Q_t(w) := \frac{1}{t} \sum_{j = 1}^t Y_j(w) ,
\end{equation}
When constructing confidence intervals at the end of a fixed time $T$, we are interested in $Q_T(w)$. However, when constructing confidence sequences, we perform inference at every time $t$; hence our estimand is the time-varying $Q_t(w)$ at every time $t$.

To connect this to the commonly known MAB reward mean distribution, consider the following example where the agent is interested in the mean reward of arm $w = 1$.
\begin{example}
\label{example:Q_func}
Suppose $t = 3$, then Equation~\eqref{eq:Q_func} reduces to
$$Q_3(1) = \frac{1}{3} (Y_1(1) + Y_2(1) + Y_3(1)).$$
\end{example}
Notice that $Q_3(1)$ in Example~\ref{example:Q_func} is the sample finite-population analogue of the more typical reward mean distribution function $E(Y_t(1))$\footnote{Here the expectation is taken with respect to the random reward.}. Because the design-based framework conditions on the full set of potential outcome, all corresponding estimands have a finite-sample interpretation for quantities of interest.



In practice, we are more interested in comparing the mean reward between two or more arms, 
\begin{equation}
\label{eq:tau}
\tau_t(w, w') := Q_t(w) - Q_t(w').
\end{equation}
For simplicity, we focus on pairwise differences between two arms, but our paper can easily be extended to estimate any linear combinations of the $K$ mean rewards. Our main result builds confidence intervals and sequences for both $Q_t(w)$ and $\tau_t(w, w')$.

\subsection{Estimation}
\label{subsection:estimates}
We now propose an estimators for $Q_t(w)$ and consequently $\tau_t(w, w')$ that serve as the main building blocks for theoretical results throughout the paper. We leverage the inverse propensity score estimator and its corresponding (conservative variance) estimator in \cite{timeseries, DB_paper}. 

First, we introduce our unbiased estimator $\hat Q_t(w)$ along with an estimate of its variance 
\begin{equation} 
\label{eq:Q_mean_var}
\begin{split}
   \hat Q_t(w) & :=  \frac{1}{t} \sum_{j = 1}^t \hat\tau_j(w) := \frac{1}{t} \sum_{j = 1}^t \frac{\mathbf{1}\{W_j = w \} Y_j}{p_{j \mid j-1}(w)} \\
    S_t(w) & :=  \sum_{j = 1}^t \hat\sigma_j^2(w) \\
    & :=  \sum_{j = 1}^t \frac{ \mathbf{1}\{W_j = w \} Y_j^2(1 - p_{j \mid j-1}(w))}{p_{j \mid j-1}(w)^2} .
\end{split}
\end{equation}

Similarly, we propose the following estimator of $\tau_t(w, w')$ and an estimate of the upper bound of its variance
\begin{equation} 
\label{eq:tau_mean_var}
\begin{split}
   \hat\tau_t(w, w') & := \hat Q_t(w) - \hat Q_t(w') = \frac{1}{t} \sum_{j = 1}^t \hat\tau_j(w) - \hat\tau_j(w') \\
    S_t(w, w') & :=  \sum_{j = 1}^t \hat\sigma_j^2(w, w')  \\
    &:= \sum_{j = 1}^t \frac{ \mathbf{1}\{W_j = w \} Y_j^2}{p_{j \mid j -1}(w)^2} + \frac{\mathbf{1}\{W_j = w' \}Y_j^2}{p_{j \mid j -1}(w')^2}.
\end{split}
\end{equation}

The above estimators are conditionally unbiased for the respective estimands, which we formally state in the following lemma that is proven in Appendix~\ref{appendix:proof_lemma}. Here, and throughout the paper, the expectation is taken with respect to the random action $W_t$.
\begin{lemma}[Unbiased properties of estimators]
\label{lemma:moment_cond}
Under Assumption~\ref{assumption:PTA}, we have that
$$E(\hat Q_t(w) \mid \mathcal{F}_{t-1}) = Q_t(w)$$
$$E(\hat\tau_t(w, w') \mid \mathcal{F}_{t-1}) = \tau_t(w, w')$$
holds for every $t= 1, \dots, T$. Furthermore, we have that
$$\text{Var}(\hat\tau_t(w) \mid \mathcal{F}_{t-1}) =  E(\hat \sigma_t^2(w)  \mid \mathcal{F}_{t-1})$$
$$\text{Var}(\hat\tau_t(w) - \hat\tau_t(w') \mid \mathcal{F}_{t-1}) \leq E(\hat\sigma_t^2(w, w')  \mid \mathcal{F}_{t-1})$$
holds for every $t = 1, \dots, T$. 
\end{lemma}

We conclude this section with a few remarks about our framework. First, besides Assumption~\ref{assumption:PTA}, which can be verified and controlled by the agent, we do not place restrictions on either the adaptive assignment process or the reward generation process. This allows us to assume that the rewards distribution is both non-$iid$ and non-stationary. Hence, all our estimands are denoted with subscript $t$ to show that these may change over time. Despite this general framework, the design-based approach allows us to perform inference since we leverage the randomness in the actions. Second, we can only obtain an upper bound of the variance for $\hat\tau_t(w) - \hat\tau_t(w')$ because the actual variance contains a product of $Y_t(w)Y_t(w')$, which is never observed. 
Finally, Lemma~\ref{lemma:moment_cond} holds for any $t$ including $t = T$. Consequently, the estimates proposed in this section are relevant for constructing both confidence intervals and confidence sequences by leveraging the martingale convergence theory. 

\section{Design-based Confidence Intervals for Multi-arm Bandits}
\label{section:CI_results}
We now demonstrate how to construct asymptotically valid confidence intervals for $Q_t(w)$ and $\tau_t(w, w')$. Before providing our result, we require an additional assumption that restricts  any realized reward to be bounded by an arbitrarily large constant. 

\begin{assumption}[Bounded realized rewards]
$$|Y_t(W_t)| \leq M$$ for all $t$ and $W_t \in \{0, \dots, K - 1\}$, where $M \in \mathbb{R}$.
\label{assumption:boundedPO}
\end{assumption}

\noindent Note that $M$ can be extreme to make this assumption hold. Such assumptions are commonly used to satisfy the necessary regularity conditions used in design-based inference \cite{timeseries, peng_bounded}. 
For example, if $T$ potential outcomes were generated from a $N(0,1)$ distribution, an unbounded distribution, each of the $T$ realized rewards are still bounded. 

\subsection{Design-based Confidence Intervals}
\label{subsection:CI}
Given these assumptions, we state the first result that allows an agent to build confidence \textit{intervals} at the end of the study at time $t = T$ for both the reward mean function $Q_T(w)$ and the reward mean difference between two arms $\tau_T(w, w')$ for general arbitrary distributions for the rewards.

\begin{theorem}[Design-based CI for MAB]
\label{theorem:CI}
Suppose data $\{w_t, y_t\}_{t = 1}^T$ are observed for a fixed pre-specified $T$, where Assumption~\ref{assumption:PTA}-~\ref{assumption:boundedPO} are satisfied and $W_t$ adapts based on the past as shown in Equation~\eqref{eq:adaptive_trt}. Then, as $T \rightarrow \infty$,
$$\hat Q_T(w) \pm  z_{1 - \alpha/2}\frac{\sqrt{\sum_{j = 1}^T \hat \sigma_j^2(w)}}{T}$$ 
forms an asymptotically valid $(1 - \alpha)$ confidence interval for $Q_T(w)$, where $z_{a}$ is the $a^\text{th}$-quantile of a standard normal distribution. Furthermore,
$$\hat\tau_T(w, w') \pm z_{1 - \alpha/2}\frac{\sqrt{\sum_{j = 1}^T \hat \sigma_j^2(w, w')}}{T}$$ 
forms an asymptotically valid $(1 - \alpha)$ confidence interval for $\tau_T(w, w')$.
\end{theorem}

The proof is provided in Appendix~\ref{appendix:proof_theom}, which uses a central limit theorem for martingale sequence differences. The widths of the confidence intervals in Theorem~\ref{theorem:CI} decrease with rate approximately $1/\sqrt{T}$, similar to that of a $t$-test. For Theorem~\ref{theorem:CI} to hold, we only required bounded realized rewards and probabilistic treatment assignments. The first is a mild condition, while the second is within the agent's control. Otherwise, we assume nothing about the data-generating process of the reward distribution and allow that the action $W_t$ is adapted based on the past. 
 
\section{Design-based Confidence Sequences for Multi-arm Bandits}
\label{section:CS_results}
We focus on the sequential nature of MAB and present a strategy to perform inference any time new data arrives by constructing valid confidence sequences. Confidence sequences are sequences of confidence intervals that are uniformly valid over time (often referred to as anytime-valid inference). Formally, a sequence set of confidence intervals $\{V_t\}_{t=1}^T$ is a valid confidence sequence with type-1 error $\alpha$ for the target parameter $\mu_t$ if
\begin{equation}
\label{eq:validity} 
\Pr(\forall t, \mu_t \in V_t) \geq 1 - \alpha 
\end{equation}
holds for arbitrary data-dependent stopping rule $T$, where $T$ is the final time of the experiment and can be determined in any data-dependent way \cite{howard_nonasymp}. In words, Equation~\eqref{eq:validity} states that our confidence sequence $V_t$ covers the desired potentially time-varying estimand estimand, \emph{e.g.}, mean reward of arm, uniformly at \textit{any time} with probability $1 - \alpha$. This formally allows the analyst to stop the MAB as soon as the analyst is satisfied with the inference, \emph{e.g.}, mean reward of an arm is statistically higher than a threshold. 

In Section~\ref{subsection:exact_CS}, we provide non-asymptotic confidence sequences that have some practical issues. Consequently, we provide an improved asymptotic confidence sequence in Section~\ref{subsection:asymp_CS}. 

\subsection{Design-Based Exact Confidence Sequence}
\label{subsection:exact_CS}
We begin by stating the exact closed-form confidence sequence in the following theorem that is proved in Appendix~\ref{appendix:closed_form}.
\begin{theorem}[Design-based Exact CS for MAB]
\label{theorem:nonasymp_CS}
Suppose data $\{w_t, y_t\}_{t = 1}^T$ are observed for any arbitrary data dependent stopping time 
$T$\footnote{With a slight abuse of notation we also use $T$ (the final data collection time) as a stopping time. Furthermore, because $T$ is data-dependent, we require that it is formally a well-defined stopping time, \emph{i.e.},  a measurable function dependent on the current and previous data (not on the future).}, where Assumption~\ref{assumption:PTA}-~\ref{assumption:boundedPO} holds. Let $m:= M/p_{min}$, where $p_{min} = \min\limits_{t, w} p_{t \mid t-1}(w)$. Then, $\hat Q_t(w)  \pm C_t(S_t(w))$ forms a valid $(1-\alpha)$ confidence sequence for $Q_t(w)$, where $C_t(S_t) := $
\begin{equation*}
\begin{scriptsize}
  \left[\frac{m(m+1)}{t} \log \Bigg( \frac{2}{\alpha} \Bigg) + \frac{S_t}{t}\left(\frac{m+1}{m}\log\Big(1 + \frac{1}{m} \Big) - \frac{1}{m} \right) \right]  
\end{scriptsize}
\end{equation*}
Furthermore, $\hat \tau_t(w, w')  \pm C_t(S_t(w, w'))$ forms a valid $(1-\alpha)$ confidence sequence for $\tau_t(w, w')$, where $S_t(w), S_t(w, w')$ are defined in Equation~\eqref{eq:Q_mean_var} and \eqref{eq:tau_mean_var}, respectively. 
\end{theorem}
There are two practical limitations to Theorem~\ref{theorem:nonasymp_CS}. First, the confidence sequence scales with $M/p_{min}$. $M$ may be unknown unless under special cases such as binary rewards. Furthermore, in the spirit of a sequential test, the agent may desire to change $p_{t \mid t-1}(w)$ as more units enter. However, Theorem~\ref{theorem:nonasymp_CS} requires $p_{min}$ to be determined before the start of the algorithm, thus restricting the action probabilities to a pre-specified range. 

Second, the confidence width is of order approximately $O(S_t/t)$, which likely only shrinks asymptotically to zero if the estimated variances grow sub-linearly or there are stronger assumptions on the potential outcomes. We can fix the second issue by leveraging a mixture distribution with the truncated gamma distribution to build another confidence sequence shown in Appendix~\ref{appendix:alternative} with order approximately $O(\sqrt{S_t \log(S_t)} /t)$. The price we pay for the additional performance gain is that the CS no longer has a closed-form solution and requires root-solving algorithms.

\subsection{Design-Based Asymptotic Confidence Sequence}
\label{subsection:asymp_CS}
We now improve the confidence sequence presented in Theorem~\ref{theorem:nonasymp_CS} by providing asymptotic confidence sequences. Informally, asymptotic confidence sequences are valid confidence sequences after a ``sufficiently large'' time. We further show the robustness of our confidence sequence through simulations in Section~\ref{section:sims}. 

For completeness, we formally define an asymptotic confidence sequence first introduced in \cite{time_uniform}.
\begin{definition}[Asymptotic Confidence Sequences]
\label{def:asymp_CS}
We say that ($\hat\mu_i \pm V_i$) is a two-sided $(1-\alpha)$ asymptotic confidence sequence for a target parameter $\mu_i$ if there exists a non-asymptotic confidence sequence $(\hat\mu_i \pm \tilde{V}_i)$ for $\mu_i$ such that 
\begin{equation}
\label{eq:asymp_CS_req}
\frac{\tilde{V}_t}{V_t} \xrightarrow{a.s.} 1.
\end{equation}
\end{definition}
To give some intuition of the above definition,  ``couplings'' has been used in the literature of strong approximations \cite{approx1, approx2} to formally define asymptotic confidence \textit{intervals}. In this literature, asymptotic confidence interval is defined by a ``coupled'' finite-sample valid confidence interval centered at the same statistic such that the difference between the two non-asymptotic and asymptotic confidence intervals is negligible in the limit. Equation~\eqref{eq:asymp_CS_req} captures the same notion except with the almost-sure convergence to satisfy the time \textit{uniform} guarantee required.\footnote{This is formally proven in Appendix C.4 of \cite{time_uniform}.} 

Before stating the theorem, we require an additional assumption that restricts the variance from vanishing in the limit.
\begin{assumption}[None Vanishing Variance]
\label{assumption:adaptive_var_no_vanish}
Let $\sigma_t^2(w) := \text{Var}(\hat\tau_t(w) \mid \mathcal{F}_{t-1})$ and $\sigma_t^2(w, w')$ is defined similarly. Then we assume that both
$$\sum_{j = 1}^t \sigma_{j}^2(w) \rightarrow \infty, \quad  \sum_{j = 1}^t \sigma_{j}^2(w, w') \rightarrow \infty$$
almost surely.
\end{assumption}
Assumption~\ref{assumption:adaptive_var_no_vanish} holds if $1/t \sum_{j = 1}^t \sigma_j^2(w) \xrightarrow{a.s.} \sigma_{*}^2$ or if $\sigma_1^2(w) = \sigma_2^2(w) = \dots = \sigma_T^2(w)$. Informally, Assumption~\ref{assumption:adaptive_var_no_vanish} is satisfied as long as the potential outcomes do not vanish to zero as time grows.

\begin{theorem}[Design-based Asymptotic CS for MAB]
Suppose data $\{w_t, y_t\}_{t = 1}^T$ are observed for any arbitrary data dependent stopping time $T$, where Assumption~\ref{assumption:PTA}-~\ref{assumption:adaptive_var_no_vanish} are satisfied and $W_t$ adapts based on the past as shown in Equation~\eqref{eq:adaptive_trt}.  Then, $\hat Q_t(w) \pm D_t(S_t(w))$ forms a valid $(1-\alpha)$ asymptotic confidence sequence for $Q_t(w)$, where
$$D_t(S_t) :=  \frac{1}{t} \sqrt{\frac{S_t \eta^2 + 1}{\eta^2} \log \Bigg( \frac{S_t \eta^2 + 1}{\alpha^2}\Bigg) } $$
forms a valid $(1-\alpha)$ asymptotic confidence sequence for $Q_t(w)$. Furthermore, $\hat\tau_t(w, w') \pm D_t(S_t(w, w'))$ forms a valid $(1-\alpha)$ asymptotic confidence sequence for $\tau_t(w, w')$, where  $\eta > 0$ is any pre-specified constant
\label{theorem:CS}
\end{theorem}
The proof is also provided in Appendix~\ref{appendix:proof_theom}. The confidence sequences provided in Theorem~\ref{theorem:CS} can also serve as valid confidence \textit{intervals} at any time $t$. The width of the confidence sequences scale similar to that in Theorem~\ref{theorem:CS} except there is an extra log term to control the type-1 error at every time uniformly. Lastly, $\eta$ is typically chosen by the analyst to minimize the confidence width at a certain fixed time. Following the advice in \cite{DB_paper}, for all examples and simulations, we choose $\eta$ so that the width is minimized at time 10. We do this because the confidence sequence width is largest at early times, and the choice of $\eta$ becomes insignificant as more data arrives. We additionally show a closed-form expression to calculate $\eta$ in Appendix~\ref{appendix:choosing_rho}. For practice, we recommend using the same $\eta$ we do throughout the paper, \emph{i.e.}, $\eta \approx 0.77$. 

To illustrate Theorems~\ref{theorem:CI} and \ref{theorem:CS}, consider the following basic example. 
\begin{figure}[ht]
\begin{center}
\includegraphics[width=12cm]{"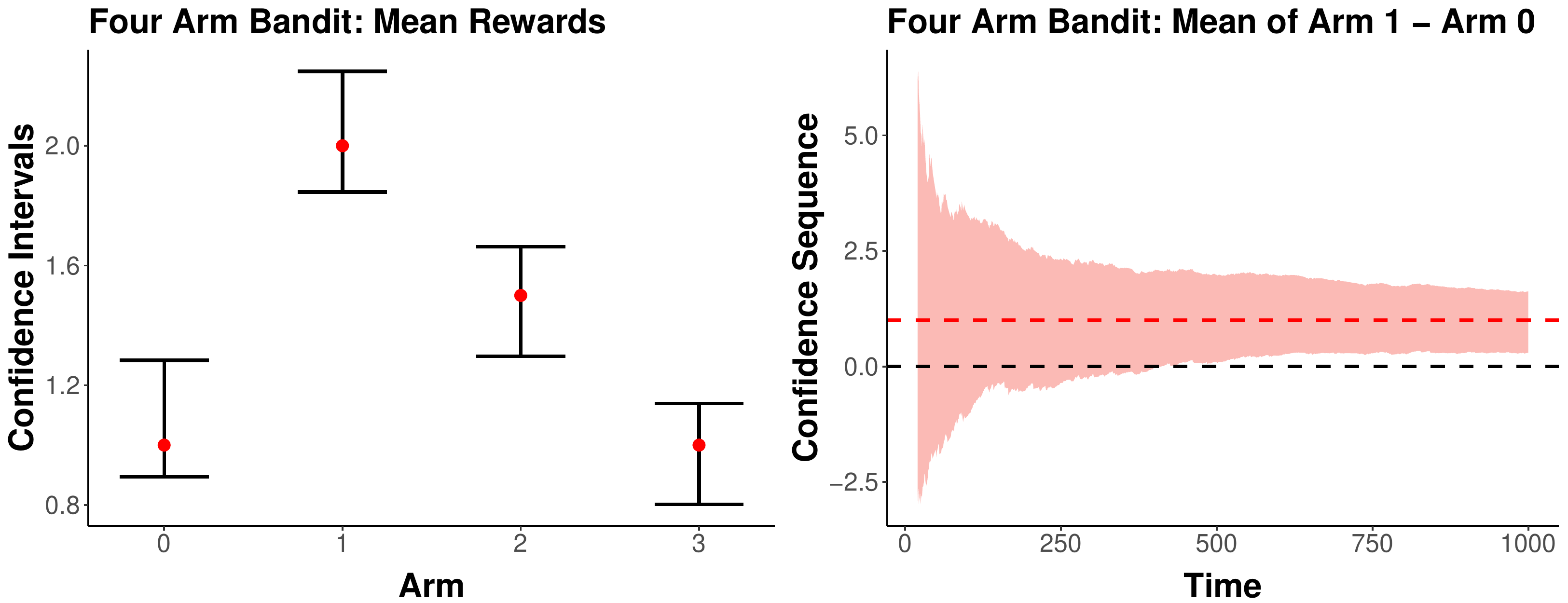"}
\caption{Four Arm Bandit (Example~\ref{example:bandit}). The left panel uses Theorem~\ref{theorem:CI} to construct confidence intervals for the four-arm means, where the red dot represents the truth. On the right panel, the red contours show the lower and upper confidence sequence using Theorem~\ref{theorem:CS}. The horizontal red dotted line represents the true mean difference of the rewards, and the black horizontal dotted line represents the zero (null) line. }
\label{fig:bandit_example}
\end{center}
\end{figure}

\begin{example}[Stationary and independent four-arm bandit]
\label{example:bandit}
Suppose a four-arm bandit problem, where an agent pulls from arms $0, 1, 2, 3$. Suppose the rewards for arm $w$ are stationary and independently generated from a $N(\mu_w, 1)$, where $\mu_0 = \mu_4 = 1, \mu_1 = 2, \mu_2 = 1.5$, so that arm 1 has the highest mean reward. 

We use the following adaptive probabilities for the $t^\text{th}$ observation 
\begin{equation}
\label{eq:adaptive_prob}
p_{t \mid t-1}(w) = \frac{\bar{Y}_{w, t-1}}{\sum_{w = 1}^K \bar{Y}_{w, t-1}}, \quad t > 0.1*T, 
\end{equation}
where $\bar{Y}_{w, t-1}$ is the sample mean of arm w using the obtained samples for up to time $t -1$. In other words, the agent up-weights the arm that produces higher mean rewards. Lastly, the agent does a fair coin flip for all the arms for the first 10\% of the sample (exploration period), where $T = 1000$. 

The left panel of Figure~\ref{fig:bandit_example} shows our proposed confidence intervals cover the true mean reward for all arms even under adaptively sampled data. Furthermore, the right panel shows our confidence sequence tightens to the desired truth and covers the true mean difference at all times. For this case, if the agent desired to find any arm that is better than the default arm 0, then this confidence sequence shows the agent can terminate the experiment as early as $t = 400$ (approximately the first time the red contours are statistically significantly above zero), saving further budget and sample. 
\end{example}

\section{Contextual multi-arm bandits}
\label{section:CMAB}
We now extend our results to contextual MAB, where an agent has access to some ``context'' or covariates $X_t$ before deciding the $t^\text{th}$ action. Consequently, at the end of the $t^\text{th}$ roud the full data are $\{X_t, W_t, Y_t\}_{t = 1}^T$, where $X_t$ may be multi-dimensional. Although we can ignore $X_t$ and only use $(W_t, Y_t)$ to construct valid confidence intervals and confidence sequences, we can leverage $X_t$ as a variance reduction technique. 

To formalize this, we first denote 
$$\hat Y_{t \mid t-1} := \hat f_{t}(X_{1:t}, Y_{1:(t-1)}, W_{1:(t-1)}),$$
where $\hat f_{t}$ denotes the prediction for the next observation $Y_t$ using all the past data and the current $X_t$. With a slight abuse of notation, we make the filtration $\mathcal{F}_{t - 1}$ to further contain the information set for $X_{1:t}$. Since our results leverage martingale theory, $\hat Y_{t \mid t-1}$ is equivalent to a constant because we always condition on $\mathcal{F}_t$. 

More formally, our estimand $\tau_t(w, w')$ in Equation~\eqref{eq:tau} can be re-written as $\tau_t(w, w') = $
\begin{equation} 
\label{eq:proxy_outcome}
\begin{split}
   & Q_t(w) - Q_t(w') \\
    &= \frac{1}{t} \sum_{j = 1}^t \{Y_j(w) -\hat Y_{j \mid j-1} \} - \{Y_j(w') -\hat Y_{j \mid j-1} \}.
\end{split}
\end{equation}
Using the above formulation, we can define our new ``residualized'' outcome $\tilde{Y}_t := Y_t - \hat Y_{t \mid t-1}$ and use the same estimators in Section~\ref{subsection:estimates} and results in Theorems~\ref{theorem:CI} and \ref{theorem:CS} except replacing $Y_t$ with $\tilde{Y}_t$. 

More formally, we rewrite Equation~\eqref{eq:tau_mean_var} as 
\begin{equation} 
\label{eq:tau_mean_var_proxy}
\begin{split}
   \hat\tau_t(w, w')^X & := \frac{1}{t} \sum_{j = 1}^t \hat\tau_j(w)^X - \hat\tau_j(w')^X \\
    S_t(w, w')^X  &:= \sum_{j = 1}^t \hat\gamma_j^2(w, w') ,
\end{split}
\end{equation}
where $\hat\tau_t(w)^X, \hat\gamma_t^2(w, w')$ are defined similar to $\hat\tau_t(w), \hat\sigma_t^2(w, w')$ except replacing $Y_t$ with $\tilde{Y}_t$. 
The result of Lemma~\ref{lemma:moment_cond} directly extends to these new estimators because we condition on the filtration; thus $\hat Y_{t \mid t-1}$ is equivalent to a constant (hence it is crucial that the predictions are constructed based on the past data without using $W_t$). This allows the analyst to formally use $\hat Y_{t \mid t-1}$ to incorporate any machine learning algorithm or prior knowledge to reduce the variance. This reduction is proportional to how small $\{Y_t - \hat Y_{t \mid t-1} \}^2$ is, \emph{i.e.},  how well the analyst can use the prior data to predict the next response. 

Furthermore, we can also allow $p_{t \mid t-1}(w)$, defined in Equation~\eqref{eq:adaptive_trt}, to further adapt based on the context variables. Note that, we can only use $\hat Y_{t \mid t-1}$ for tackling $\tau_t(w, w')$ since $\hat Y_{t \mid t-1}$ does not cancel in Equation~\eqref{eq:proxy_outcome} when just estimating $Q_t(w)$ alone. 

\subsection{Variance Reduction Technique}
We now extend Theorems~\ref{theorem:CI} and \ref{theorem:CS} for building valid confidence intervals and sequences for $\tau_t(w, w')$ using $\hat Y_{t \mid t-1}$ in the contextual MAB setting. 

\begin{theorem}[Design-based CI and CS for contextual MAB]
\label{theorem:CI_CMAB}
Suppose data $\{x_t, w_t, y_t\}_{t = 1}^T$ are observed, where Assumption~\ref{assumption:PTA}-~\ref{assumption:boundedPO} are satisfied and $W_t$ adapts based on the past as shown in Equation~\eqref{eq:adaptive_trt}. Further suppose $\hat Y_{t \mid t-1}$ is bounded for all $t$. Then, 
$$\hat\tau_T(w, w')^X \pm z_{1 - \alpha/2}\frac{\sqrt{\sum_{j = 1}^T \hat\gamma_j^2(w, w')}}{T}$$ 
forms an asymptotically valid $(1 - \alpha)$ confidence interval for $\tau_T(w, w')$ for a pre-specified $T$.

Denote $\tilde{\sigma}_j^2(w, w') := \text{Var}(\hat\tau_t(w ,w)^X \mid \mathcal{F}_{t-1})$. Then if we further assume none vanishing variance for the new residualized outcome, \emph{i.e.},  $\sum_{j = 1}^t \tilde{\sigma}_j^2(w, w') \rightarrow \infty$ almost surely, then 
$$\hat\tau_t(w, w')^X \pm D_t(S_t(w, w')^X)$$
forms a valid $(1-\alpha)$ asymptotic confidence sequence for $\tau_t(w, w')$ for any arbitrary data-dependent stopping rule $T$, where  $\eta > 0$ is any pre-specified constant and $D_t(S_t)$ is defined in Theorem~\ref{theorem:CS}.
\end{theorem}

The proof is omitted because under the stated assumptions, the setting is identical to that of Theorems~\ref{theorem:CI} and \ref{theorem:CS} except we replace the response $Y_t$ with the new residualized response $\tilde{Y}_t$. We show through simulations in Section~\ref{subsection:sim_cont} that residualizing the outcome can lead to a substantial reduction in the variance.

\section{Simulations and Related Work}
\label{section:sims}
We now provide a simulation with two goals. First, we show the empirical coverage of the proposed methods and compare the advantages and disadvantages of using confidence sequences over confidence intervals in a simple $iid$ binary reward setting. Next, we shift to a more complex setting with non-stationary continuous rewards with context variable $X_t$. We demonstrate both the validity and gain we gain from incorporating $X_t$ into our confidence interval and sequence. Finally, we end with a discussion of our results in context with existing works. 

\subsection{Independent and Identically Distributed Binary Rewards}
\label{subsection:sim_bin}
We begin our simulations in a simplistic setting where we have two arms $w = 0, 1$ with $iid$ binary rewards from $\text{Bern}(\mu_w)$, where $\mu_1 = 0.27, \mu_0 = 0.15$, \emph{i.e.}, a 12\% expected increase from choosing arm 1. Although this is a simplified setting, it represents typical adaptive A/B tests or MAB with a treatment and control group, \emph{e.g.}, learning whether the new product is better than the standard offering through bandits. 

For simulations in Section~\ref{subsection:sim_bin}-~\ref{subsection:sim_cont}, we build confidence interval and confidence sequence for $\tau_t(1, 0)$, \emph{i.e.}, the mean reward difference from choosing arm 1 over arm 0 when $\alpha = 0.05$. For simplicity, we use the adaptive sampling procedure outlined in Equation~\eqref{eq:adaptive_prob}, where we use the first 10\% of samples for exploration and adapt based on the sample means of each arm. Additionally, we run 1000 Monte-Carlo experiments and report four statistics. We first report the coverage, \emph{i.e.}, the proportion of times the confidence interval covers $\tau_T(1, 0)$ for a fixed $T$. Then, for confidence sequences, we report the proportion of times our confidence sequence covers $\tau_t(1, 0)$ for all times $t > 10$. Following the advice in \cite{DB_paper}, we check only after an initial 10 samples because our method is asymptotic, and it is practically unlikely for the analyst to terminate the experiment only after 10 samples. Second, we report the average width at a pre-specified time $t = T$ for all methods. Third, we report the average stopping time for only the confidence sequences, where our stopping time is defined as the first time the confidence sequence does not cover zero. Therefore, for all confidence sequences, we run the simulation for sufficiently large $t > T$. Lastly, we report the statistical power for the confidence interval.

Table~\ref{tab:sim_simple} shows the simulation results under the simple $iid$ setting for $T = 700$ samples. We find that all our proposed methods have the desired coverage. As expected, the width for the confidence sequences is wider than that of the confidence interval by less than two times. However, on average, the asymptotic confidence sequence can detect an effect as early as $t = 580$ (approximately 80\% of $T = 700$ the total sample). Although the analyst would reject close to 92\% of the times with the proposed confidence interval by $t = 700$, the confidence sequences are attractive alternatives if the agent wishes to terminate the experiment as soon as the agent detects a statistically significant effect. Lastly, we find that the asymptotic confidence sequence has improved width and stopping time compared to the non-asymptotic confidence sequence while maintaining proper coverage at even early times, likely due to our conservative variance estimator. 

\begin{table}[t]
\begin{center}
\begin{adjustbox}{max width=\textwidth,center}
\label{tab:sim_simple}
\begin{tabular}{|l|c|c|c|c|}
 \hline
Method & Coverage & Width & Stopping Time & Power \\ 
 \hline
Asymp-CI  & 95\% & 0.14 & NA & 92\% \\
 \hline
Asymp-CS  & 95\% & 0.23 & 580 & NA \\
 \hline
Exact-CS & 99\% & 0.25 & 640 & NA\\
 \hline
CS with X & 98\% & 1.92 & 115 & NA \\
 \hline
\end{tabular}
\end{adjustbox}
\caption{Binary $iid$ rewards simulation. The first row shows the performance of the asymptotic CI in Theorem~\ref{theorem:CI} while the second and third row shows the performance of the asymptotic and exact CS using Theorem~\ref{theorem:CS} and Theorem~\ref{theorem:nonasymp_CS}, respectively under $\alpha = 0.05$.}
\end{center}
\end{table}

\subsection{Non-stationary Continuous Rewards}
\label{subsection:sim_cont}
We now change our reward distribution to a non-stationary continuous distribution with four arms ($K = 4$). We further add one binary context variable $X_t$ to illustrate Theorem~\ref{theorem:CI_CMAB}. More specifically, our data generating process is the following AR(1) linear model
\begin{equation} 
\label{eq:sim_DGP}
\begin{split}
    Y_t(0) &= \rho Y_{t-1}(0) + \beta X_t + \epsilon_{t},|\rho| \leq 1, \epsilon_{t} \overset{iid}{\sim} N(\mu_{t, 0}, 1) \\
    Y_{t}(w) &= Y_{t}(0) + N(\mu_{t, w}, 1) \\
    Y_{0}(0) &= 0, \quad  X_t \sim \text{Bern}(0.5),
\end{split}
\end{equation}
where $\rho$ represents how the next potential outcome is lag-1 dependent on its immediate history. Furthermore, $\mu_{t, 1} - \mu_{t, 0}$ is our target parameter, \emph{i.e.},  the mean causal effect of being in arm 1 over arm 0. Although we can make $\mu_{t, w}$ time-varying, we fix $\mu_{t, 0} = 1, \mu_{t, 1} = 1.5, \mu_{t, 2} = 1.25, \mu_{t, 3} = 1$ for all $t$ for simplicity. Our reward distribution is non-stationary and no longer $iid$ due to $\rho$, which we set at $\rho = 0.1$. Finally, we let $\beta = 1$ and use a linear regression of $Y_{1:(t-1)}$ on $X_{1:(t-1)}$ to predict $Y_t$ to demonstrate Theorem~\ref{theorem:CI_CMAB}.

\begin{table}[t]
\begin{center}
\begin{adjustbox}{max width=\textwidth,center}
\label{tab:sim_cont}
\begin{tabular}{|l|c|c|c|c|}
 \hline
Method & Coverage & Width & Stopping Time & Power \\ 
 \hline
CI no X & 95\% & 1.84 & NA & 88\% \\
 \hline
CS no X  & 99\% & 3.68 & 340 & NA \\
 \hline
CI with X & 95\% & 1.02 & NA & 98\%\\
 \hline
CS with X & 98\% & 1.92 & 115 & NA \\
 \hline
\end{tabular}
\end{adjustbox}
\caption{Continuous non-stationary rewards simulation. The first two row shows the performance of the asymptotic CI and CS in Theorem~\ref{theorem:CI} and Theorem~\ref{theorem:CS}, respectively. The last two rows show the corresponding CI and CS incorporating the context variable $X_t$ through Theorem~\ref{theorem:CI_CMAB}. }
\end{center}
\end{table}

Table~\ref{tab:sim_cont} shows the simulation results under the non-stationary setting described in Equation~\eqref{eq:sim_DGP} for $T = 300$ samples. As expected, our methods have an over-conservative coverage due to the conservative variance estimator. Nevertheless, incorporating $X$ using Theorem~\ref{theorem:CI_CMAB} successfully reduces the stopping time by a third, approximately halving the width of the confidence sequence and interval, and increasing the power substantially. Furthermore, we similarly find that the confidence sequence width is larger than that of the confidence intervals but has the potential to end the MAB substantially earlier than $T = 300$. 

\subsection{Comparison with Related Works}
\label{subsection:related_works}
We now extend the previous simulation to compare with existing work. Our confidence interval result is most closely related to \cite{hadad}, where the aforementioned paper takes a super-population approach to inference, \emph{i.e.},  the potential outcomes $Y_t(W_t)$ are generated $iid$ from a distribution satisfying technical conditions, \emph{e.g.}, bounded moments. Additionally, to the best of our knowledge, \cite{ian_bandit, bandit_CS} are the only existing works that build confidence sequence for MAB. However, the main results presented in the aforementioned papers assume bounded $[0, 1]$ stationary rewards. Furthermore, there also exist many technical and often untestable assumptions on the rewards, which our work bypasses through the design-based approach. 

Therefore, we only compare our method with \cite{hadad}. In particular, we use Theorem 2 in \cite{hadad} with unity weights since our adaptive allocation probabilities do not diverge or oscillate asymptotically. Since \cite{hadad} takes a super-population approach, the corresponding estimand is $E(Y(0))$, where the expectation is taken with respect to the stochastic potential outcomes. 

\begin{figure}[ht]
\begin{center}
\includegraphics[width=7cm]{"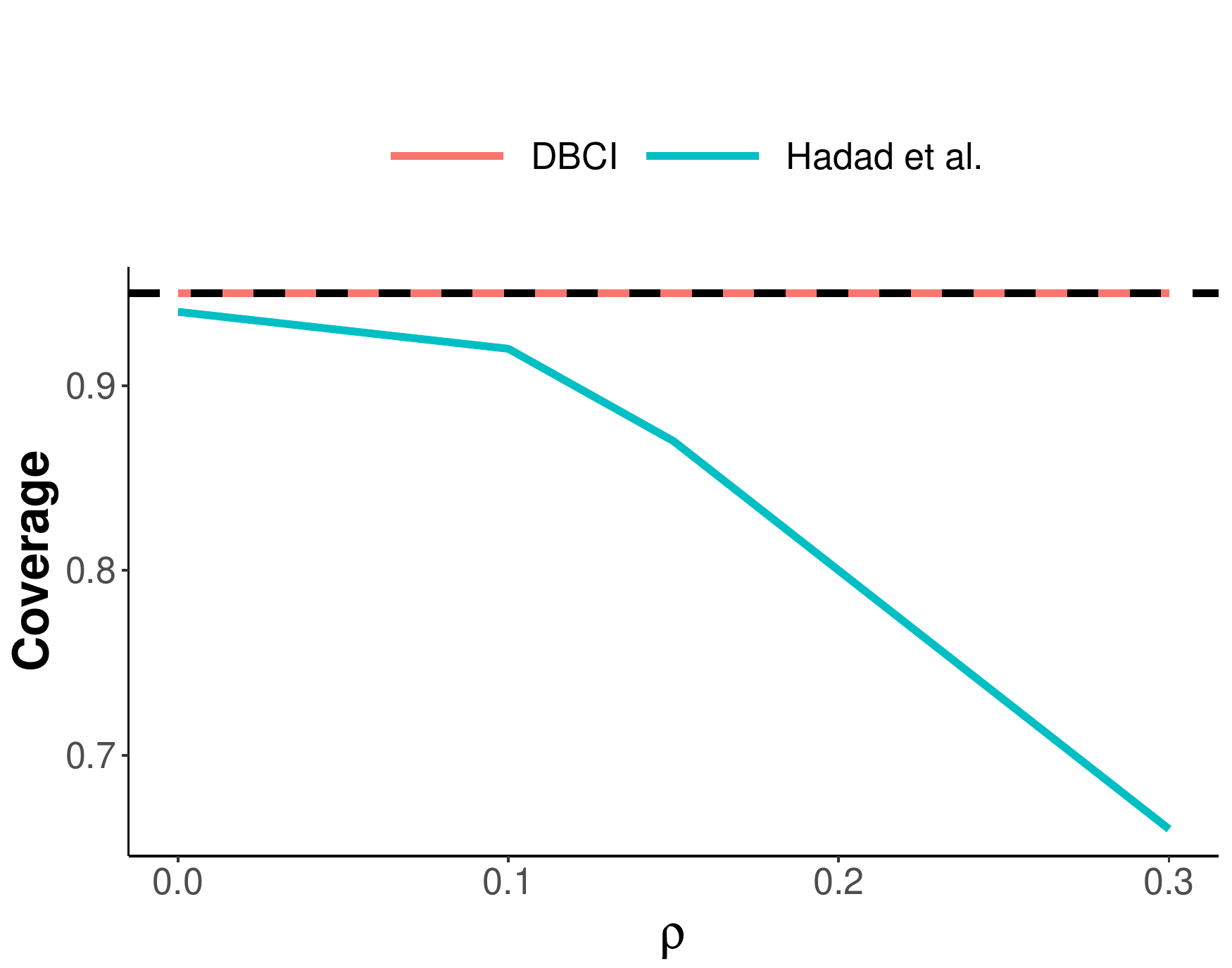"}
\caption{The simulation study settings remains identical to that in Section~\ref{subsection:sim_cont} except we vary $\rho$ to break stationarity and set $\beta = 0$. The red line shows the coverage of $Q_T(0)$ using Theorem~\ref{theorem:CI}. The blue line shows the coverage of $E(Y(0))$ using the method proposed in \cite{hadad}, respectively. Finally, the black dotted line shows the desired type-1 error at $\alpha = 0.05$.}
\label{fig:comparison}
\end{center}
\end{figure}

We keep an identical simulation setting of that in Section~\ref{subsection:sim_cont} except we set $\beta = 0$ to simplify the setting and vary $\rho$ in the $x$-axis to induce dependency and break the $iid$ stationarity assumption. We also focus on estimating $Q_T(0)$, i.e., the mean reward for arm 0. We remark that $E(Y(0)) = \mu_{t, 0}$ marginally, which is the target estimand for \cite{hadad}. Since $\rho$ induces dependency among the potential outcomes, we expect poor type-1 error control in \cite{hadad} while our proposed confidence interval in Theorem~\ref{theorem:CI} remains valid regardless of any potential outcome distribution. Figure~\ref{fig:comparison} shows that the coverage for the design-based confidence interval (DBCS) remains at the desired 95\% level while the confidence interval in \cite{hadad} quickly loses validity as $\rho$ grows. Therefore, our work extends existing work for inference in multi-arm bandits to general non-stationary MAB in an assumption-light manner through the design-based approach.

\bibliography{example_paper} 
\bibliographystyle{pa}

\newpage
\appendix

\section{Proof of Lemma~\ref{lemma:moment_cond}}
\label{appendix:proof_lemma}
\begin{align*}
  E(\hat\tau_t(w) \mid \mathcal{F}_{t-1}) &= E\left(\frac{\mathbf{1}\{W_t = w \} Y_t}{p_{t \mid t-1}(w)} \right) \\
 &= \frac{p_{t \mid t-1}(w) Y_t(w) }{p_{t \mid t-1}(w)}  \\
 &= Y_t(w)
\end{align*}
Hence, from above we immediately have that the first conditional moments are indeed unbiased. 

Next, we first calculate the closed form expression of $\text{Var}(\hat\tau_t(w) \mid \mathcal{F}_{t-1})$.
\begin{align*}
    \text{Var}(\hat\tau_t(w) \mid \mathcal{F}_{t-1}) &= \text{Var}\left(\frac{\mathbf{1}\{W_t = w \} Y_t}{p_{t \mid t-1}(w)}  \mid \mathcal{F}_{t- 1} \right) \\
    &= \frac{p_{t \mid t-1}(w)(1 - p_{t \mid t-1}(w))Y_t(w)^2}{p_{t \mid t-1}(w)^2} \\
    &= \frac{(1 - p_{t \mid t-1}(w))Y_t(w)^2}{p_{t \mid t-1}(w)} 
\end{align*}
Therefore, it follows that $\text{Var}(\hat\tau_t(w) \mid \mathcal{F}_{t-1}) =  E(\hat \sigma_t^2(w)  \mid \mathcal{F}_{t-1})$

Finally, we calculate the closed form expression of $\text{Var}(\hat\tau_t(w) - \hat\tau_t(w') \mid \mathcal{F}_{t-1})$.
\begin{align*}
    \text{Var}(\hat\tau_t(w) - \hat\tau_t(w') \mid \mathcal{F}_{t-1}) 
    &= \text{Var}(\hat\tau_t(w)  \mid \mathcal{F}_{t-1})  + \text{Var}(\hat\tau(w') \mid \mathcal{F}_{t-1}) \\
    & - 2\text{Cov}(\hat\tau_t(w), \hat\tau_t(w')  \mid \mathcal{F}_{t-1}).
\end{align*}
From the previous result we can calculate the two variance terms. First, we have that 
$$\text{Cov}(\mathbf{1}\{W_t = w \}, \mathbf{1}\{W_t = w' \} \mid \mathcal{F}_{t-1}) = - p_{t \mid t-1}(w) p_{t \mid t-1}(w').$$
Putting this together, we have that $\text{Var}(\hat\tau_t(w) - \hat\tau_t(w') \mid \mathcal{F}_{t-1}) =$
\begin{align*}
    &= \frac{(1 - p_{t \mid t-1}(w))Y_t(w)^2}{p_{t \mid t-1}(w)} + \frac{(1 - p_{t \mid t-1}(w'))Y_t(w')^2}{p_{t \mid t-1}(w')}  \\
    &+ 2Y_t(w)Y_t(w')  \\
    & \leq \frac{\left(p_{t \mid t-1}(w)  Y_t(w') + p_{t \mid t-1}(w)  Y_t(w') \right)^2}{p_{t \mid t-1}(w)p_{t \mid t-1}(w')} \\
    & \leq \frac{ Y_t(w)^2}{p_{t \mid t -1}(w)} +\frac{ Y_t(w')^2}{p_{t \mid t -1}(w')}
\end{align*}
where the third line follows because $p_{t \mid t-1}(w') \leq (1 - p_{t \mid t-1}(w))$ for any $w, w'$ and the last line follows because $(a - b)^2 \geq 0 \rightarrow a^2 + b^2 \geq 2ab$ and we let $a = p_{t \mid t-1}(w')  Y_t(w)$ and $b = p_{t \mid t-1}(w)  Y_t(w')$. Finally we see that it is straight forward to show that 
$$E(\hat\sigma_t^2(w, w')  \mid \mathcal{F}_{t-1}) = \frac{ Y_t(w)^2}{p_{t \mid t -1}(w)} +\frac{ Y_t(w')^2}{p_{t \mid t -1}(w')},$$
completing the proof. 

\section{Proof of Theorems~\ref{theorem:CI} and \ref{theorem:CS}}
\label{appendix:proof_theom}
Throughout this proof, we omit subscript ``a.s.'' from $o_{a.s.}(.)$ and $O_{a.s.}(.)$ to simplify notation. The proof proceeds in several steps, where the first step is enough to prove Theorem~\ref{theorem:CI}. The proof is identical for proving valid confidence interval and confidence sequence for either $Q_t(w)$ or $\hat\tau_t(w, w')$. Therefore, we show the proof for $\hat\tau_t(w, w')$.  

\paragraph{Step 1: Strong Approximation via Martingale Sequence Differences}
The following is identical to Step 2 of the proof in Appendix D \cite{DB_paper}, but we provide it here for completeness. 

We first define
$$u_t = \hat\tau_{t}(w, w') - \tau_{t}(w, w').$$ 
By Lemma~\ref{lemma:moment_cond}, $\{u_t\}$ is a martingale difference sequence with respect to $F_{T, t-1}$. Similar to the proof of Step 2 of \cite{time_uniform}, we also use the strong approximation theorem presented in \cite{stratssen}. In particular, we require Equation (159) in Theorem 4.4 of Strassen's paper (further details in Lemma A.3 of \cite{time_uniform}) for our strong approximation theorem. However, our proof is different than that in Step 2 of \cite{time_uniform} for the following reason. 

The original Theorem 4.4 in \cite{stratssen} is stated for martingales difference sequence of the form $E(X_n \mid \sigma(X_1, \dots, X_{n-1})) = 0$, where we use Strassen's notation and $X_i$ are random variables with defined second moment. Although our martingale is of the form $E(f(X_n) \mid \sigma(X_1, \dots, X_{n-1})) = 0$, where $f(.)$ is the function that maps the data to $u_t$. More formally, to use the strong approximation theorem in \cite{stratssen}, we replace the beginning conditions of Theorem 4.4 in the following way. 

``Let $X_1, X_2, \dots$ be random variables such that $0 \leq E(f(X_n)^2 \mid X_1, \dots, X_{n-1}) \leq C$ is bounded by some constant $C$ (this directly holds under Assumption~\ref{assumption:boundedPO}) and $E(f(X_n) \mid X_1, \dots, X_{n-1}) = 0$, a.s. for all $n$. Put $S_n = \sum_{i \leq n} f(X_i)$ and $V_n = \sum_{i \leq n} E(f(X_i)^2 \mid X_1, \dots, X_{i -1})$, where, in order to avoid trivial complications, we assume $V_1 = E(f(X_1)^2) > 0$.'' 

The remaining conditions are identical and we omit the uniform integrability condition in Equation (138) of \cite{stratssen} since it holds trivially under our bounded potential outcome for Assumption~\ref{assumption:boundedPO}. We remark that the proof leading to Equation (159) remains identical and valid except replacing $X_n$ with $f(X_n)$ in the appropriate steps. In particular, all random variables are still measurable with respect to $\sigma(X_1, \dots, X_{n-1})$. Lastly, although this theorem uses the actual variance (not an upper bound), using an upper bound only makes the confidence sequence width strictly wider and hence the validity still holds.

Let $Z_1, \dots, Z_t$ be $iid$ random standard Gaussian. Utilizing Theorem 4.4 Equation (159) in \cite{stratssen} we have that 
\begin{equation}
\label{eq:strong_approx}
    \frac{1}{t} \sum_{j = 1}^t u_j = \frac{1}{t} \sum_{j = 1}^t \sigma_{j}(w, w') Z_j + o\left(\frac{\tilde{S}_{t}^{3/8}\log(\tilde{S}_{t})}{t} \right) \quad a.s.,
\end{equation}
where $\tilde{S}_t := \sum_{j = 1}^t \sigma_j(w, w')^2$. Simplifying Equation~\eqref{eq:strong_approx}, we have that 
$$\frac{(\hat\tau_t(w, w') - \tau_t(w, w'))}{\sqrt{1/t^2 \sum_{j = 1}^t \sigma_j(w, w')^2}} \approx N(0, 1).$$
Our final result in Theorem~\ref{theorem:CI} follows directly after replacing a conservative or exact consistent estimator for the variance $\sigma_j(w, w')$ with $\hat\sigma_j(w, w')$ (akin to the difference between $z$-test and $t$-test). We show that we indeed get a consistent estimator in the last step of this proof. The remainder of the proof aims to prove Theorem~\ref{theorem:CS}.

\paragraph{Step 2: Building martingale using Gaussian distribution}
Recently, \citep{martingale_must} shows that all sequential tests must have an explicit or implicit construction of a non-negative martingale. Although one of the major advantages of an asymptotic confidence sequences is that it avoids explicitly constructing a martingale, the proof still relies on constructing a martingale with the asymptotic Gaussian distribution. Consequently, the first step of the proof builds a martingale from a sequence of $iid$ standard Gaussian random variables. 

We note that
$$M_t(\lambda):= \text{exp}\left( \sum_{j = 1}^t (\lambda \sigma_{j}(w, w') Z_j - \lambda^2 \sigma_{j}(w, w')^2/2) \right)$$
is a non-negative martingale starting at one for any $\lambda \in \mathbb{R}$ with respect to the canonical filtration \citep{filtration_cite}. For algebraic simplicity, we also define $L_t:= \sum_{j = 1}^t \sigma_{j}(w, w') Z_j$ and $\bar{\sigma}_t^2 = \frac{1}{t} \sum_{j = 1}^t \sigma_{j}(w, w')^2$. Moreover, for any probability distribution $F(\lambda)$ on $\mathbb{R}$, we also have the mixture, 
$$\int_{\lambda \in \mathbb{R}} M_t(\lambda) dF(\lambda) $$
is again a non-negative martingale with initial value one \citep{filtration_cite}. In particular, we consider the probability distribution function $f(\lambda; 0, \eta^2)$ for the Gaussian distribution with mean zero and variance $\eta^2$ as the mixing distribution. The resulting martingale is
\begin{align*}
    M_t:=&  \int_{\lambda \in \mathbb{R}} M_t(\lambda) f(\lambda; 0, \eta^2) d\lambda \\
    =& \frac{1}{\sqrt{2\pi\eta^2}} \int_{\lambda} \text{exp}\left(\lambda  L_t - \frac{t\lambda^2 \bar{\sigma}_t^2 }{2}  \right) \text{exp}\left( \frac{-\lambda^2}{2\eta^2}\right) d\lambda \\
    =& \frac{1}{\sqrt{2\pi\eta^2}} \int_{\lambda} \text{exp}\left(\lambda  L_t - \frac{\lambda^2(1 + t\eta^2 \bar{\sigma}_t^2)}{2\eta^2} \right) d\lambda \\
    =& \frac{1}{\sqrt{2\pi\eta^2}} \int_{\lambda} \text{exp}\left(\frac{-\lambda^2(1 + t\eta^2 \bar{\sigma}_t^2 ) + 2 \lambda \eta^2 L_t} {2\eta^2}  \right) d\lambda \\
    =& \frac{1}{\sqrt{2\pi\eta^2}} \int_{\lambda} \text{exp}\left(\frac{-a(\lambda^2 + \frac{b}{a} 2\lambda} {2\eta^2}  \right) d\lambda ,
\end{align*}
where $a = t\eta^2\bar{\sigma}_t^2 + 1$ and $b = \eta^2 L_t$. Completing the square, we have that the integrand is:
$$\text{exp}\left(\frac{-a(\lambda^2 + \frac{b}{a} 2\lambda} {2\eta^2}  \right)  =  \text{exp}\left(\frac{-(\lambda - b/a)^2}{2\eta^2/a}\right) \text{exp}\left(\frac{b^2}{2a\eta^2}\right).$$
Putting the expression back into $M_t$ we have that,
\begin{align*}
M_t &= \frac{1}{\sqrt{2\pi\eta^2/a}}   \int_{\lambda} \text{exp}\left(\frac{-(\lambda - b/a)^2}{2\eta^2/a}\right) d\lambda \frac{\text{exp}\left(\frac{b^2}{2a\eta^2}\right)}{\sqrt{a}} \\
&= \text{exp}\left( \frac{\eta^2(\sum_{j=1}^t \sigma_{j}(w, w') Z_j)^2}{2(t \bar{\sigma}_t^2 \eta^2 + 1)}\right) (t \bar{\sigma}_t^2 \eta^2 + 1)^{-1/2},
\end{align*}
\noindent where the last line follows because the first part of the first line is one and we plug back in the definition of $a$ and $b$.

Since $M_t$ is a non-negative martingale with initial value one we can use Ville's maximal inequality \cite{vile} to claim that 
$$\Pr(\forall t \geq 1, M_t < 1/\alpha) \geq 1 - \alpha,$$
which simplifies to the following after taking the logarithm and simple algebraic manipulation.
\begin{equation}
\label{eq:gaussian_CS}
\Pr\left(\forall t \geq 1, \left|\frac{1}{t} L_t \right| < \sqrt{ \frac{2(t\bar{\sigma}_t^2 \eta^2 + 1)}{t^2 \eta^2} \text{log}\Bigg( \frac{\sqrt{t\bar{\sigma}_t^2\eta^2 + 1}}{\alpha} \Bigg)} \right) \geq 1- \alpha 
\end{equation}

Combining Equation~\eqref{eq:gaussian_CS} and Equation~\eqref{eq:strong_approx} implies that with probability at least $(1 - \alpha)$, 
\begin{equation}
\label{eq:nonasymp_proofCS}
  \forall t \geq 1, \left|\frac{1}{t} \sum_{i=1}^t u_i \right| < \sqrt{ \frac{2(t\bar{\sigma}_t^2 \eta^2 + 1)}{t^2 \eta^2} \text{log}\Bigg( \frac{\sqrt{t\bar{\sigma}_t^2\eta^2 + 1}}{\alpha} \Bigg)} +  o\left(R_t\right),
\end{equation}
where $R_t = \tilde{S}_{t}^{3/8}\log(\tilde{S}_{t})/t$. Using Assumption~\ref{assumption:adaptive_var_no_vanish}, we have that
\begin{equation}
\hat\tau_t(w, w') \pm \sqrt{ \frac{2(t\bar{\sigma}_t^2 \eta^2 + 1)}{t^2 \eta^2} \text{log}\Bigg( \frac{\sqrt{t\bar{\sigma}_t^2\eta^2 + 1}}{\alpha} \Bigg)}
\label{eq:true_CS}
\end{equation}
forms an $(1 - \alpha)-$asymptotic confidence sequence, where we used Assumption~\ref{assumption:adaptive_var_no_vanish} so that the $\tilde{V}_t/V_t \xrightarrow{a.s.} 1$ holds where $\tilde{V}_t$ is the non-asymptotic confidence width in Equation~\eqref{eq:nonasymp_proofCS} (with the little $o$ term) and $V_t$ is defined in Equation~\eqref{eq:true_CS} (without the little $o$ term). 

\paragraph{Step 3: Using empirical variance} Unfortunately, the confidence sequence in Equation~\eqref{eq:true_CS} can not be directly used because $\bar{\sigma}_t$ is based off the true variance and hence not obtainable from the data. The last step is to replace Equation~\eqref{eq:true_CS} with our estimated variance.

First, if $1/t \sum_{j = 1}^t \hat\sigma_j(w, w')^2 \xrightarrow{a.s.} \bar{\sigma}_t^2$ (strongly consistent variance estimate), then we first show our confidence sequence in Theorem~\ref{theorem:CS}
forms a $(1-\alpha)$-asymptotic confidence sequence for $\tau_t(w, w')$, giving us the desired result. 

We begin by rewrite the assumption of $1/t \sum_{j = 1}^t \hat\sigma_j(w, w')^2 \xrightarrow{a.s.} \bar{\sigma}_t^2$ as $\tilde{\sigma}_t^2 - \bar{\sigma}_t^2 = o(\bar{\sigma}_t^2)$, where $\tilde{\sigma}_t^2 = 1/t \sum_{j = 1}^t \hat\sigma_j(w, w')^2$. Then, we have that our width from Equation~\eqref{eq:true_CS} 
\begin{align*}
 & \sqrt{ \frac{2(t\bar{\sigma}_t^2 \eta^2 + 1)}{t^2 \eta^2} \text{log}\Bigg( \frac{\sqrt{t\bar{\sigma}_t^2\eta^2 + 1}}{\alpha} \Bigg)} \\
 &= \sqrt{ \frac{2(t (\tilde{\sigma}_t^2 + o(\bar{\sigma}_t^2)) \eta^2 + 1)}{t^2 \eta^2} \text{log}\Bigg( \frac{\sqrt{t (\tilde{\sigma}_t^2 + o(\bar{\sigma}_t^2))\eta^2 + 1}}{\alpha} \Bigg)}  \\
 &= \sqrt{ \frac{t (\tilde{\sigma}_t^2 + o(\bar{\sigma}_t^2) ) \eta^2 + 1}{t^2 \eta^2} \text{log}\Bigg( \frac{t (\tilde{\sigma}_t^2 + o(\bar{\sigma}_t^2))\eta^2 + 1}{\alpha^2} \Bigg)} \\
 &= \sqrt{ \frac{t\tilde{\sigma}_t^2\eta^2 + o(t \bar{\sigma}_t^2)+ 1}{t^2 \eta^2} \text{log}\Bigg( \frac{t \tilde{\sigma}_t^2\eta^2 + o(t\bar{\sigma}_t^2) + 1}{\alpha^2} \Bigg)} \\
 &= \sqrt{ \left( \frac{t\tilde{\sigma}_t^2\eta^2 + 1}{t^2 \eta^2} + o(\bar{\sigma}_t^2/t) \right) \text{log}\Bigg( \frac{t \tilde{\sigma}_t^2\eta^2 + o(t\bar{\sigma}_t^2) + 1}{\alpha^2} \Bigg)} \\
\end{align*}
Focusing on the second logarithmic term, we have
\begin{align*}
\log \Bigg( \frac{t \tilde{\sigma}_t^2\eta^2 + o(t\bar{\sigma}_t^2) + 1}{\alpha^2} \Bigg) &= \text{log}\Bigg( \frac{t \tilde{\sigma}_t^2\eta^2 + 1}{\alpha^2} + o(t\bar{\sigma}_t^2) \Bigg)\\
&= \text{log}\Bigg( \frac{t \tilde{\sigma}_t^2\eta^2 + 1}{\alpha^2} \big[1 + o(1) \big]  \Bigg)\\
&= \text{log}\Bigg( \frac{t \tilde{\sigma}_t^2\eta^2 + 1}{\alpha^2}  \Bigg) + \log(1 + o(1)) \\
&= \text{log}\Bigg( \frac{t \tilde{\sigma}_t^2\eta^2 + 1}{\alpha^2}  \Bigg) + o(1),
\end{align*}
where the last line follows because $\text{log}(1 + x) = x + o(1)$ for $|x| < 1$. Returning back to the main expression we have  
\begin{align*}
 &\sqrt{ \frac{2(t\bar{\sigma}_t^2 \eta^2 + 1)}{t^2 \eta^2} \text{log}\Bigg( \frac{\sqrt{t\bar{\sigma}_t^2\eta^2 + 1}}{\alpha} \Bigg)} \\
 &= \sqrt{ \left( \frac{t\tilde{\sigma}_t^2\eta^2 + 1}{t^2 \eta^2} + o(V_t  /t^2) \right) \left[ \text{log}\Bigg( \frac{t \tilde{\sigma}_t^2\eta^2 + 1}{\alpha^2}  \Bigg)  + o(1) \right] } \\
 &= \sqrt{ \frac{t\tilde{\sigma}_t^2\eta^2 + 1}{t^2 \eta^2} \text{log}\Bigg( \frac{t \tilde{\sigma}_t^2\eta^2 + 1}{\alpha^2} \Bigg) + o(V_t/t^2) + o(V_t \log V_t/t^2 ) + o(V_t/t^2) } \\
  &= \sqrt{ \frac{t\tilde{\sigma}_t^2\eta^2 + 1}{t^2 \eta^2} \text{log}\Bigg( \frac{t \tilde{\sigma}_t^2\eta^2 + 1}{\alpha^2} \Bigg) + o(V_t \log V_t/t^2 )  } \\
 &= \sqrt{ \frac{t\tilde{\sigma}_t^2\eta^2 + 1}{t^2 \eta^2} \text{log}\Bigg( \frac{t \tilde{\sigma}_t^2\eta^2 + 1}{\alpha^2} \Bigg)} + o(\sqrt{V_t \log V_t}/t) ,
\end{align*}
where the last line follows because $\sqrt{a + b} \leq \sqrt{a} + \sqrt{b}$ for $a, b \geq 0$. This formally shows how our confidence sequence in Theorem~\ref{theorem:CS} is a valid $(1 - \alpha)-$asymptotic confidence sequence for given that our variance estimator is strongly consistent. 

However, Lemma~\ref{lemma:moment_cond} only tells us that $\tilde{\sigma}_t^2$ is conditionally unbiased for $\bar{\sigma}_t^2$. To establish the consistency result, we again use a version of strong law of large numbers for martingale sequence difference. We denote $U_t:=  \tilde{\sigma}_t^2 - \bar{\sigma}_t^2$. We remark that $U_t$ is a martingale sequence difference with respect to the filtration $F_{T, t-1}$. Using classical results in \citep{chow_conv}, we have that $U_t \xrightarrow{a.s.} 0$ since Assumption~\ref{assumption:boundedPO} immediately satisfies the needed uniformly integrability condition. Since all the convergence statements above are almost-sure convergence, steps 1-3 give the desired claim. 

\section{Exact confidence sequences}
\label{appendix:exact_CS}

\subsection{Proof of Closed-form Exact Confidence Sequence}
\label{appendix:closed_form}
\begin{proof}
We build off the proof of Theorem 4 in \citep{howard_nonasymp}. For brevity, we prove the theorem only for constructing valid confidence sequences for $\tau_t(w, w')$ and denote $S_t$ for $S_t(w, w')$ for brevity. We first show that 
$$\exp\left[\frac{\sum_{j = 1}^t(\hat\tau_j(w, w') - \tau_j(w, w'))}{m(m+1)}  + \frac{S_t}{m^2} \left(\log\Big(\frac{m}{m+1} \Big) + \frac{1}{m+1} \right) \right]$$
is a non-negative supermartingale with respect to the filtration $\mathcal{F}_{t- 1}$. \cite{fan_lemma} shows that 
$$\exp\left( \lambda\kappa  + \kappa^2 (\lambda + \log(1 - \lambda)) \right) \leq 1 + \lambda \kappa$$
for $\kappa \geq -1$ and $\lambda \in [0, 1)$. We let 
$$\kappa = \frac{\hat\tau_t(w, w')}{m},$$
where $\kappa \geq -1$ since $|\hat \tau_t(w, w')| \leq m$ for every $t$ by Assumption~\ref{assumption:boundedPO}. Let $h(\lambda) := (\lambda + \log(1 - \lambda))$ we have 
\begin{align*}
& \exp\left( \lambda \frac{ \hat\tau_t(w, w')}{m}  +  \frac{\hat\tau_t(w, w')^2}{m^2}h(\lambda)  \right) \leq 1 + \lambda \frac{ \hat\tau_t(w, w')}{m} \\
& E \left[\exp\left( \frac{\lambda \hat\tau_t(w, w')}{m}  + \frac{\hat\sigma_t^2(w, w')}{m^2} h(\lambda) \right) \mid \mathcal{F}_{t-1} \right]  \leq 1 + \lambda \frac{\tau_t(w, w')}{m} \\
& E \left[\exp\left( \frac{\lambda (\hat\tau_t(w, w') - \tau_t(w, w'))}{m}  + \frac{\hat\sigma_t^2(w, w')}{m^2} h(\lambda) \right) \mid \mathcal{F}_{t-1} \right]  \leq 1
\end{align*}
where the second line follows because $\hat\tau_t(w, w')^2 = \hat\sigma_t^2(w, w')$ and Lemma~\ref{lemma:moment_cond} and the last line follows because $1 - x \leq \exp(-x)$. We plug $\lambda = 1/(m+1)$ and because the above is a non-negative quantity this directly implies that 
$$\exp\left[\frac{\sum_{j = 1}^t(\hat\tau_j(w, w') - \tau_j(w, w'))}{m(m+1)} + \frac{S_t}{m^2} \left(\log\Big(\frac{m}{m+1} \Big) + \frac{1}{m+1} \right) \right]$$
is indeed a non-negative super martingale with respect to $\mathcal{F}_{t-1}$ as desired with initial value less than one. Therefore, we apply Ville's maximal inequality \cite{vile} to obtain
\begin{align*}
 & \Pr\left(\exists t: \exp\left[\frac{\sum_{j = 1}^t u_j}{m(m+1)}  + \frac{S_n}{m^2} \left(\log\Big(\frac{m}{m+1} \Big) + \frac{1}{m+1} \right) \right] \geq \frac{1}{\tilde{\alpha}} \right)  \\
&= \Pr\left(\exists t: \left[\frac{\sum_{j = 1}^t u_j}{m(m+1)}  + \frac{S_n}{m^2} \left(\log\Big(\frac{m}{m+1} \Big) + \frac{1}{m+1} \right) \right] \geq \log \Bigg( \frac{1}{\tilde{\alpha}} \Bigg) \right)  \\
&= \Pr\left(\exists t: \sum_{j = 1}^t u_j \geq m(m+1) \log \Bigg( \frac{1}{\tilde{\alpha}} \Bigg) -  \frac{(m+1)S_n}{m} \left(\log\Big(\frac{m}{m+1} \Big) + \frac{1}{m+1} \right)  \right) \\
&= \Pr\left(\exists t: \sum_{j = 1}^t u_j \geq \left[m(m+1) \log \Bigg( \frac{1}{\tilde{\alpha}} \Bigg) + S_n \left(\frac{m+1}{m}\log\Big(1 + \frac{1}{m} \Big) - \frac{1}{m} \right) \right] \right) 
\end{align*}
is less than $\tilde{\alpha}$. This gives the one-sided confidence sequence and we can do the same trick and build the same statement instead for $\kappa = -\hat\tau_t(w, w')/m$. Taking $\alpha = \tilde{\alpha}/2$ and applying the union bound completes the proof. 
\end{proof}

\subsection{Alternative exact confidence sequence}
\label{appendix:alternative}
In this section, we correct the order of the confidence sequence presented above. We leverage the results presented in \citep{ian_exact_extension, howard_nonasymp} by applying a mixture martingale over a truncated gamma distribution. 

The above proof shows that
\begin{equation}
\label{eq:super_martingale}
M_t:= \exp\left( \lambda A_t  + B_t (\lambda + \log(1 - \lambda)) \right) 
\end{equation}
is a super-martingale with initial value 1, where 
$$A_t:= \frac{\sum_{j = 1}^t u_j }  {m}, \quad B_t:= \frac{S_t}{m^2}.$$
For any distribution $F$ on $(0,1)$, we have by Fubini's theorem that 
$$\tilde{M}_t:= \int_{\lambda \in (0, 1)} M_t dF(\lambda) $$
is again another super-martingale with initial value 1. Following the proof of Theorem 2 in \citep{ian_exact_extension}, we choose the truncated gamma distribution given by
$$f(\lambda) = \frac{\rho^\rho e^{-\rho \left(1 - \lambda\right)} \left(1 - \lambda\right)^{\rho - 1}}{\Gamma(\rho) - \Gamma(\rho, \rho)}$$
for any $\rho \geq 0$. Therefore, we have that 
\begin{align*}
\tilde{M}_t &= \int_0^1 \exp \left \{ \lambda A_t + B_t(\lambda + \log(1 - \lambda) \right \} f(\lambda) d\lambda \\
&=\int_0^1 \exp \left \{ \lambda A_t + B_t(\lambda + \log(1 - \lambda) \right \} \frac{\rho^\rho e^{-\rho \left(1 - \lambda\right)} \left(1 - \lambda\right)^{\rho - 1}}{\Gamma(\rho) - \Gamma(\rho, \rho)} d\lambda \\
&= \frac{\rho^\rho e^{-\rho}}{\Gamma(\rho) - \Gamma(\rho, \rho)} \int_0^1 \exp \{\lambda \left(A_t + B_t + \rho \right)\} \left(1 - \lambda\right)^{B_t + \rho - 1} d\lambda \\
&= \left(\frac{\rho^\rho e^{-\rho}}{\Gamma(\rho) - \Gamma(\rho, \rho)}\right) \left(\frac{1}{B_t + \rho}\right) {_1F_1}(1, B_t + \rho + 1, A_t + B_t + \rho),
\end{align*}
where the last line follows from the definition of the Kummer's confluent hypergeometric function. 

Therefore, we have Ville's maximal inequality \cite{vile} that 
$$\Pr\left(\exists t: \left(\frac{\rho^\rho e^{-\rho}}{\Gamma(\rho) - \Gamma(\rho, \rho)}\right) \left(\frac{1}{B_t + \rho}\right) {_1F_1}(1, B_t + \rho + 1, A_t + B_t + \rho) \geq \frac{1}{\tilde{\alpha}} \right)$$
is less than $\alpha$. Consequently, a one-sided lower confidence sequence can be obtained by a root-finding algorithm to find all 
$$\{\tau_t(w, w'): V_t(w, w') \geq \frac{1}{\tilde{\alpha}} \},$$
where 
$$V_t(w, w'):= \left(\frac{\rho^\rho e^{-\rho}}{\Gamma(\rho) - \Gamma(\rho, \rho)}\right) \left(\frac{1}{B_t + \rho}\right) {_1F_1}(1, B_t + \rho + 1, A_t + B_t + \rho).$$
An upper confidence sequence can be obtained in a similar way. We remark that this confidence sequence does not solve the issue where it requires the analyst to know $M$ and $p_{min}$ before the experiment. Furthermore, this confidence sequence does not have a closed-form expression, thus it requires a root-solving algorithm to build the confidence sequence. However, \citep{ian_exact_extension} show that this provably has an asymptotic rate of $O(\sqrt{B_t \log(B_t)}/t)$, which does solve the issue related to the order of the confidence sequence width.

\section{Optimizing and choosing hyper-parameter}
\label{appendix:choosing_rho}
In this section, we show in detail how an analyst can choose $\eta$ to optimize the confidence sequence width for a desired specific time $t^*$. We remark that the derivations are nearly identical to those presented in \cite{time_uniform}, but we repeat them here for completeness. 

Our proposed confidence sequence width presented in this paper all have the following structure
$$B_t(\alpha):= \sqrt{ \frac{2(t \eta^2 + 1)}{t^2\eta^2} \text{log}\Bigg( \frac{\sqrt{t \eta^2 + 1}}{\alpha}\Bigg) }, $$
where we have omitted the variance terms and instead substituted each 1 since we want $\eta$ to be data-independent.\footnote{Consequently, we are not formally optimizing $\eta$ for the actual confidence width, but $\eta$ can still be conceptually interpreted as minimizing the confidence sequence width at a desired time $t^*$ (See Appendix C.3 in \cite{time_uniform} for more details).} We remark that
$$\argmin_{\eta > 0} B_t(\alpha) = \sqrt{\argmin_{x > 0} f(x)},$$
$$\text{where } \quad f(x):= \frac{t  x + 1}{t^2 x} \text{log}\left(\frac{t x + 1}{\alpha^2} \right), \quad x:= \eta^2.$$
Furthermore, $\lim_{x \rightarrow 0} f(x) = \lim_{x \rightarrow \infty} f(x)$ and thus if we can find the critical point by finding a solution for $\partial{f}/\partial{x} = 0$, then this must be the unique minimum.

Therefore, we have that 
$$\frac{\partial{f}}{\partial{x}} = -\frac{1}{t^2 x^2}\text{log}\left(\frac{tx + 1}{\alpha^2} \right) + \frac{1}{tx}.$$
Setting the above to zero, we obtain
$$-\alpha^2 \text{exp}(1) = -(tx + 1) \text{exp}(-(tx + 1)).$$
Therefore, we have that the solution is $-(tx + 1) = W_{-1}(-\alpha^2 \text{exp}(1))$, where $W_{-1}$ is the lower branch of the Lambert $W$ function. The solution only exists if 
$$-\alpha^2 \text{exp}(1) \geq -\text{exp}(1),$$
or equivalently if $\alpha^2 \leq 1$, which is always true for any $\alpha \in [0, 1]$. Therefore, we have that 
$$\argmin_{\eta > 0} B_t(\alpha) = \sqrt{\frac{-W_{-1}(-\alpha^2 \text{exp}(1)) - 1}{t^* }}.$$

\end{document}